\documentclass[aps,preprintnumbers,superscriptaddress,twocolumn,amsmath,amssymb,floatfix,prl]{revtex4} 
\pdfoutput=1
\usepackage{graphicx}
\usepackage[usenames,dvipsnames]{color}
\usepackage[colorlinks,bookmarks=false,citecolor=NavyBlue,linkcolor=Red,urlcolor=blue]{hyperref}
\usepackage[export]{adjustbox}

\newcommand\be{\begin{equation}}
\newcommand\bea{\begin{eqnarray}}
\newcommand\bes{\begin{subequations}}
\newcommand\esu{\end{subequations}}
\newcommand\ee{\end{equation}}
\newcommand\eea{\end{eqnarray}}

\newcommand{\cmmnt}[1]{}

\def\doi{http://dx.doi.org/}

\usepackage{braket}

\newcommand \dd{{\rm d}}

\begin{document}

\title{Exact local correlations and full counting statistics for arbitrary states of the one-dimensional interacting Bose gas}

\author{Alvise Bastianello}
\affiliation{SISSA \& INFN, via Bonomea 265, 34136 Trieste, Italy}
\author{Lorenzo Piroli}
\affiliation{SISSA \& INFN, via Bonomea 265, 34136 Trieste, Italy}
\author{Pasquale Calabrese}
\affiliation{SISSA \& INFN, via Bonomea 265, 34136 Trieste, Italy}


\begin{abstract}
We derive exact analytic expressions for the $n$-body local correlations in the one-dimensional Bose gas with contact repulsive 
interactions (Lieb-Liniger model) in the thermodynamic limit. 
Our results are valid for arbitrary states of the model, including ground and thermal states, stationary states after a quantum quench, 
and non-equilibrium steady states arising in transport settings. 
Calculations for  these states are explicitly presented and physical consequences are critically discussed.
We also show that the $n$-body local correlations are directly related to the full counting statistics for the particle-number fluctuations in  
a short interval, for which we provide an explicit analytic result.

\end{abstract}


\maketitle

\emph{Introduction--}
During the last decade, a series of ground breaking experiments with one-dimensional (1D) ultracold gases \cite{BlDZ08,CCGO11,PSSV11,KiWW04,LOHP04,PWMM04,KiWW05,kinoshita-2006,AEWK08,AJKB10,JABK11,hrm-11,FPCF15,PMCL14} opened the way 
to reveal many-body quantum effects that were unaccessible before. 
Indeed, 1D many-body systems are special for a twofold reason: on the one hand, quantum effects are enhanced 
in low dimension \cite{giamarchi-04}; on the other hand, they can be theoretically treated with powerful analytic tools such as integrability \cite{gaudin,BaxterBook,TakahashiBook,MussardoBook,guan2013,KorepinBook}
and conformal field theory \cite{giamarchi-04,GNTbook,cft-book,Card08}.
Let us e.g. consider a  dilute 1D gas of $N$ bosons of mass $m$, which will be the main focus of this work. 
In standard experimental conditions, the interaction between atoms can be approximated 
by a contact interaction \cite{Olsh98} and the  Hamiltonian in second quantization is
\be
H=\frac{\hbar^2}{2m} \int_{0}^{L}\,{\rm d}x 
\left[\partial_x\psi^{\dagger}(x)\partial_x\psi(x)+\kappa \left(\psi^{\dagger}(x)\right)^2\left(\psi(x)\right)^2\right]\,,
\label{eq:hamiltonian}
\ee
where $\psi^{\dagger}$ and $\psi$ are canonical bosonic operators and $\kappa$ is an interaction parameter. This Hamiltonian corresponds to the celebrated Lieb-Liniger (LL)  model \cite{LiLi63}, one of the most studied integrable systems. 

Although the LL model is integrable and we have exact analytic predictions for different thermodynamic quantities \cite{YaYa69,AEWK08},
the experimentally measured  correlation functions \cite{KiWW05,AJKB10,JABK11,hrm-11,FPCF15} are not easily calculated and in most of the cases we are restricted to either numerical methods \cite{ag-03,ScFl07,dbsp-07,VeCi10} or hybrid numerical-analytic techniques \cite{CaCa06,ZWKG15} (see however \cite{pk-13}). This is true also in the simplest case of one-point functions
 such as the $n$-body local correlators
\be
g_n\equiv\frac{\langle [\psi^\dagger(x)]^n[\psi(x)]^n \rangle}{D^n},
\label{eq:def_gn}
\ee
where $D\equiv N/L$ is the density of the gas.

In this Letter we consider the longstanding problem of the exact computation of $g_n$ and we provide close analytical expressions  valid for every $n$ and in an arbitrary excited state.
Besides being measurable for different values of $n$ \cite{KiWW05,AJKB10,LOHP04,JABK11,hrm-11}, the local correlators $g_n$
provide fundamental information about the system \cite{BGMH97,bz-11,odml-17}, such as particle losses \cite{BGMH97} 
and, consequently, about the overall stability of the gas. 
As a further result, we unveil another interesting application, namely the relation of $g_n$ with the 
\emph{full counting statistics} for the particle-number fluctuations within an interval \cite{HLSI08,KPIS10,KISD11,GKLK12}.

The one-point functions \eqref{eq:def_gn} are arguably the simplest correlations to be considered. Still, the problem of their computation has puzzled physicists for more than an entire decade. It is important to note that, while the interest was initially limited to thermal states \cite{JiMi81,OlDu03,GaSh03,KGDS03,NRTG16,PiCa16}, more recently increasing attention was devoted to the case of arbitrary macrostates of the system \cite{KoMT09,Pozs11mean,KoCI11,Pozs11,DWBC14,PiCa15,PiCE16}, such as generalized Gibbs ensembles (GGEs) \cite{RDYO07,ViRi16}. 
Furthermore, our result can be combined with generalized hydrodynamics \cite{CaDY16,BCDF16} to access  
 the profiles of $[\psi^\dagger(x)]^k[\psi(x)]^k$ in inhomogeneous systems \cite{InPreparation}.

The two-body local correlation $g_2$ is easily obtained for any excited state of the Hamiltonian \eqref{eq:hamiltonian}, using the Hellmann-Feynman theorem and the knowledge of exact thermodynamics \cite{GaSh03,KGDS03}. Conversely the three-body correlator $g_3$ is extremely hard to work out, and even at zero temperature its computation was initially performed only through a theoretical tour de force \cite{ChSZ06}. 
A major breakthrough was the rewriting of the LL Hamiltonian as the non-relativistic (NR) limit of  the sinh-Gordon (ShG) field 
theory \cite{KoMT09},
which provided an analytic expression of $g_3$, valid for arbitrary macrostates \cite{KoCI11}. 
Later, following an alternative approach \cite{SBGK07}, multiple integral representations for $g_n$ were derived in \cite{Pozs11}. Despite their conceptual importance, these expressions are too complicated for actual evaluation for $n>4$.

The NR limit introduced in \cite{KoMT09}, and later exploited in \cite{KoCI11}, requires as a starting point the one-point correlation function in the ShG field theory. Until recently, the only way to approach this crucial problem has been through the LeClair-Mussardo conjecture \cite{LeMu99} (now proven in \cite{Pozs11mean}) which provides a \emph{form factor expansion} \cite{Smirnov92}, consisting in a series of multiple integrals. Although this sum is known to be quickly convergent in several relevant cases, 
only the first few terms can be evaluated, relegating its applicability to the (important) limit of small excitation density. 
In this respect, the results in the LL model derived from the NR limit of the ShG field theory suffer from the same limitations. 

Very recently, a closed expression for the expectation values of vertex operators in the ShG model has been derived for thermal states by Negro and Smirnov in \cite{NeSm13}; this result was later further simplified in \cite{BePC16}, where it was also argued to be valid for arbitrary macrostates (including GGEs). While not rigorously proven, it has been explicitly checked against the LeClair-Mussardo expansion \cite{NeSm13,BePC16} and then numerically verified in the semiclassical limit \cite{BDWY17}. The goal of this Letter is to use the Negro-Smirnov formula as input for the non-relativistic limit.

Two main results will be presented in this Letter. The first one is an explicit expression for the local $n$-body correlations \eqref{eq:def_gn}, which is valid for arbitrary macrostates of the model. The derivation of our final formula is long and will be only sketched in this Letter; the most technical part of the derivation, which is suited for a more specialized audience, will be presented elsewhere \cite{InPreparation}. The second result is a formula for the full counting statistics of the density operator in a small interval, which is entirely expressed in terms of the correlations \eqref{eq:def_gn}. Altogether, our findings constitute a remarkable example where the full counting statistics can be worked out explicitly in the presence of interaction and for arbitrary states.

\emph{The model---} We consider the LL model \eqref{eq:hamiltonian} with periodic boundary conditions; we focus on the repulsive regime $\kappa>0$ and work in dimensionless units $\hbar=2m=1$. The Bethe ansatz approach \cite{KorepinBook} provides a complete set of eigenstates of $H$; they are identified by sets of distinct real {\it rapidities} $\{\lambda_i\}_{i=1}^N$, which generalize the concept of single-particle momenta for a free system. In the thermodynamic limit, rapidities associated with a given eigenstate arrange themselves according to a given distribution function $\rho(\lambda)$. Within the framework of the thermodynamic Bethe ansatz (TBA) \cite{TakahashiBook}, one also introduces a distribution of holes, $\rho_h(\lambda)$, which is analogous to the concept routinely employed in the description of free Fermi gases. Due to interaction, the relation between $\rho(\lambda)$ and $\rho_h(\lambda)$ is non-trivial, but encoded in the thermodynamic Bethe equation
\be
\rho(\lambda)+\rho_h(\lambda)=\frac{1}{2\pi}+\int_{-\infty}^{+\infty}\frac{{\rm d}\mu}{2\pi}\,\varphi(\lambda-\mu)\rho(\mu)\,,
\label{eq:thermo_bethe}
\ee
with $\varphi(\lambda)=2\kappa/(\lambda^2+\kappa^2)$. 

In the following, it is convenient to also introduce the filling function $\vartheta(\lambda)=\rho(\lambda)/(\rho(\lambda)+\rho_h(\lambda))$; together with Eq.~\eqref{eq:thermo_bethe}, the latter uniquely specifies a macrostate. In this work, we will consider ground and thermal states \cite{TakahashiBook}, as well as stationary states reached after the system is brought out of equilibrium by a quantum quench \cite{cem-16}; these correspond to GGEs, whose rapidity distributions have been explicitly computed in a few relevant cases \cite{KSCC13,DWBC14,Bucc16}. More generally, our results are valid for arbitrary states, displaying an extremely large number of applications, from the non-equilibrium steady states arising in transport \cite{CaDY16,BCDF16} to split Fermi 
seas \cite{kinoshita-2006,KoCI11,fec-14}.

\begin{figure*}
	\begin{center}
		\includegraphics[width=0.32\textwidth,valign=l]{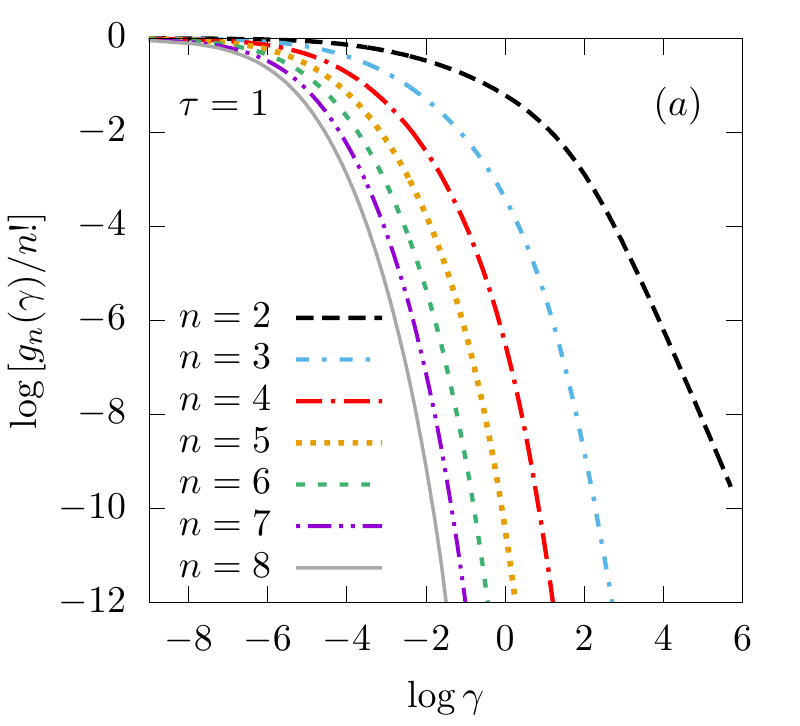}
		\includegraphics[width=0.32\textwidth,valign=l]{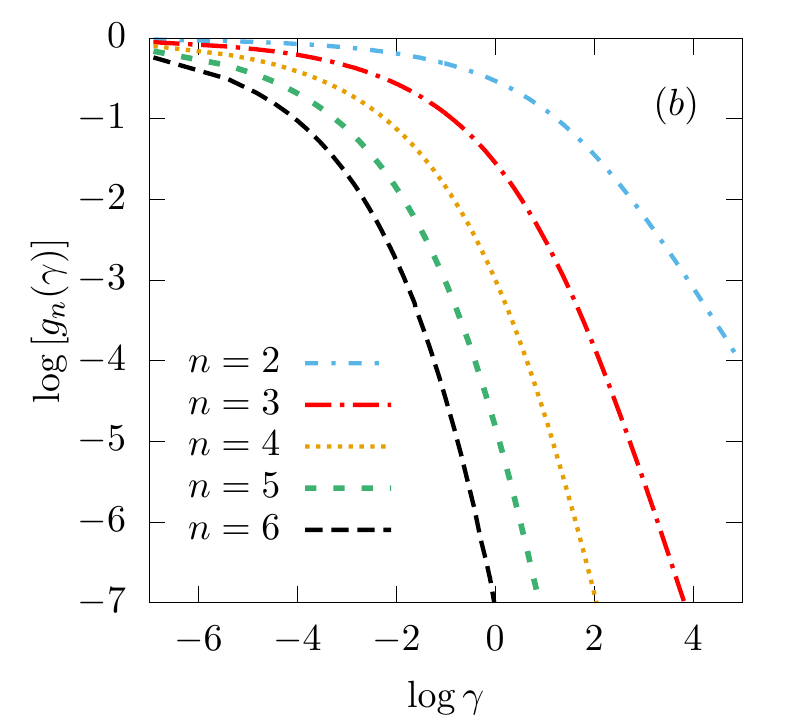}
            \includegraphics[width=0.32\textwidth,valign=l]{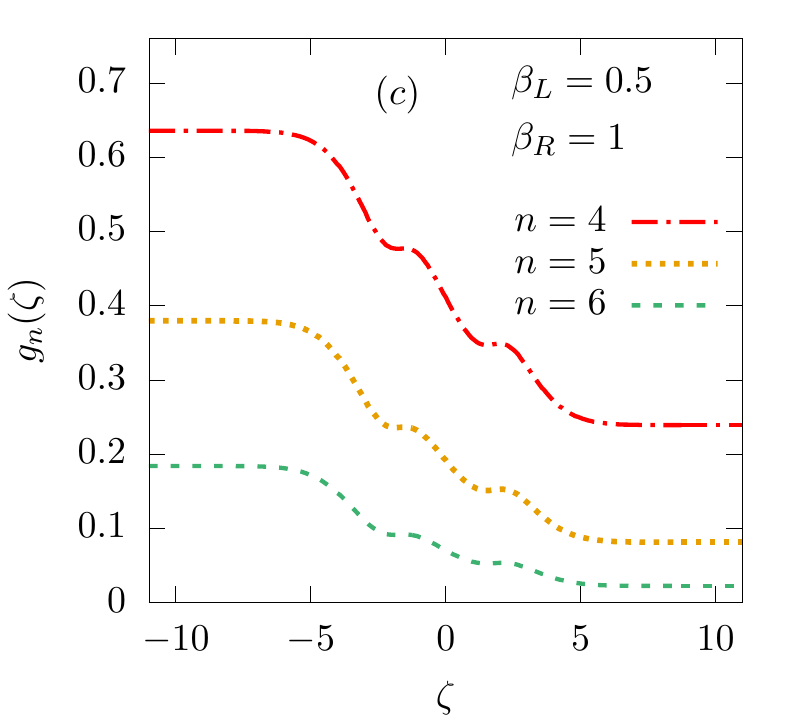}
	
\caption{Correlators $g_n$ for different values of $n$. 
		($a$): $g_n(\gamma)$ in a thermal state with $\tau=1$. We report the rescaled value $g_n(\gamma)/n!$. ($b$): $g_n(\gamma)$ in the steady state after the quench studied in \cite{DWBC14}. ($c$): Profiles of $g_n(\zeta)$ as functions of the ray $\zeta=x/t$ in the nonhomogenous protocol of \cite{BCDF16,CaDY16}: the two halves of the system here are initialized in two thermal states with $\tau_L=2$, $\tau_R=1$, $D_L=D_R=1$ and $\kappa=0.5$.}
		\label{fig:thermal_quench}
	\end{center}
\end{figure*}

\emph{Analytic formula for $g_n$---} Our main result for the local $n$-body function is
\be
g_n=\frac{(n!)^2\kappa^n}{2^nD^n}
\sum_{\sum_j jm_j=n}\prod_\ell \left[\frac{1}{m_\ell!}\left(\frac{\mathcal{B}_\ell}{\pi\kappa} \right)^{m_\ell}\right]\,.
\label{eq:main_corr}
\ee
Here the sum is over all sequences of positive integers $\{m_j\}_{j}$ s.t. $\sum_{j}jm_j=n$, 
the constants $\mathcal{B}_n=n^{-1}\int \dd \lambda\, \vartheta(\lambda)b_{2n-1}(\lambda)$ are 
functions of $b_n(\lambda)$, obtained as the solution to the system
\bea
b_{2n}(\lambda)&=&\int \frac{\dd\mu}{2\pi}\,\left\{ \varphi(\lambda-\mu)\vartheta(\mu)[b_{2n}(\mu)-b_{2n-2}(\mu)]\right.\nonumber\\
&+&\left.\Gamma(\lambda-\mu)\vartheta(\mu)(2b_{2n-1}(\mu)-b_{2n-3}(\mu))\right.\,, 
\label{eq:int_1}
\eea
\bea
b_{2n+1}(\lambda)&=&\delta_{n,0}+\int \frac{\dd\mu}{2\pi}\, \big\{\,\Gamma(\lambda-\mu)\vartheta(\mu) b_{2n}(\mu)\nonumber\\
&+&\, \varphi(\lambda-\mu)\vartheta(\mu)[b_{2n+1}(\mu)-b_{2n-1}(\mu)]\big\}\,,
\label{eq:int_2}
\eea
where $b_{n\le 0}(\lambda)\equiv 0$, and $\Gamma(\lambda)=\lambda \varphi(\lambda)/\kappa$.
Finally $\vartheta(\lambda)$ is the filling function characterizing the macrostate. 
We now discuss the main physical consequences of this result and postpone its derivation to the end of the Letter.
For $n=2,3,4$, Eq. \eqref{eq:main_corr} is a representation alternative  to those in \cite{KoCI11,Pozs11}; 
we  checked numerically the equivalence between the two, as well as perturbatively in the functional 
parameter $\vartheta(\lambda)$ \cite{InPreparation}; furthermore, we also numerically verified that our formulas for the thermal case are consistent with the analytic expansions derived in \cite{NRTG16}, for large values of $\gamma$.

The system of equations in \eqref{eq:int_1} and \eqref{eq:int_2} is linear, and the solution for each function $b_{n}(\lambda)$ 
involves only a finite number of equations; accordingly, the numerical implementation presents no difficulty. 
As an important example, we explicitly evaluated Eq.~\eqref{eq:main_corr} for different macrostates reported 
in Fig.~\ref{fig:thermal_quench}. 
In subfigure $(a)$ we report explicit values of the correlations for a thermal state $\hat{\rho}=e^{-\beta H}/{\rm tr}\left[e^{-\beta H}\right]$. 
In this case, the correlators $g_n$ are only a function of the rescaled parameters $\gamma=\kappa/D$ and $\tau=\beta^{-1}D^{-2}$. 
As expected, the value of $g_n$ decreases for increasing $\gamma$, and we verified that it algebraically vanishes for $\gamma\to\infty$.
Furthermore, in the free limit we recover the expected behavior $g_n(\gamma=0)=n!$. 
It is interesting to compare these qualitative features with those of the post-quench steady state after a quantum quench. 
We consider the protocol of \cite{DWBC14}, in which the system is initially prepared in the ground-state of the non-interacting Hamiltonian,
and the interaction suddenly turned on to a finite value $\kappa$. 
The results also for this case are reported in Fig. \ref{fig:thermal_quench}, panel ($b$), with limiting behavior
$g_n(\gamma=0)=1$ and with an algebraical decay for large $\gamma$, generalizing the results of \cite{DWBC14} for $n=2$, $3$. Finally, we apply our formula to the inhomogeneous out-of-equilibrium protocol studied in \cite{CaDY16,BCDF16}: two different halves of the system are prepared in different macrostates, which are joined together at the origin and subsequently left to evolve with the Hamiltonian \eqref{eq:hamiltonian}.
At late times, time- and space-dependent local quasi-stationary states emerge \cite{BeFa16}, and relaxation to a GGE occurs at each ``ray'' $\zeta=x/t$, where $x$ is the distance from the junction \cite{CaDY16,BCDF16};  accordingly, each observable displays a nontrivial profile as a function of the rescaled variable $\zeta$. This is illustrated in subfigure $(c)$ of Fig.~\ref{fig:thermal_quench}, where the profiles of $g_n$ obtained after joining together two different thermal states are displayed.

\emph{The full counting statistics---} 
We now report on the relation of the local correlators with the full counting statistics of the particle density on a small interval.
Together with Eq.~\eqref{eq:main_corr}, this constitutes the main result of this Letter. 
The probability distribution of observables within a finite length interval recently became the subject of intensive theoretical 
investigations \cite{gadp-06,cd-07,lp-08,sfr-11,cmv-12,ia-13,sk-13,k-14,lla-15,lddz-15,sp-17,nr-17,hb-17,CoEG17} boosted by  several experimental measurements with cold atoms  \cite{HLSI08,KPIS10,KISD11, GKLK12,AJKB10,JABK11}. 
Nevertheless, no exact analytic results for interacting integrable models were previously known.

We consider the probability $P_\Delta(n)$ of having $n$ particles in an interval of length $\Delta$ centered around $x$. 
It holds
\be
\lim_{\Delta\to 0} \frac{P_\Delta(n)}{\Delta^n}=\frac{\langle [\psi^\dagger(x)]^n[\psi(x)]^n \rangle}{n!}\, ,
\label{eq:prob}
\ee
providing the leading contribution to $P_\Delta(n)$ in the limit of small intervals. 
For a finite interval, Eq.~\eqref{eq:prob} is a good approximation provided $\Delta$ is smaller than two length scales, 
as it will be clear from the derivation below. 
\begin{figure}[b]
	\begin{center}
		\includegraphics[width=1\columnwidth,valign=l]{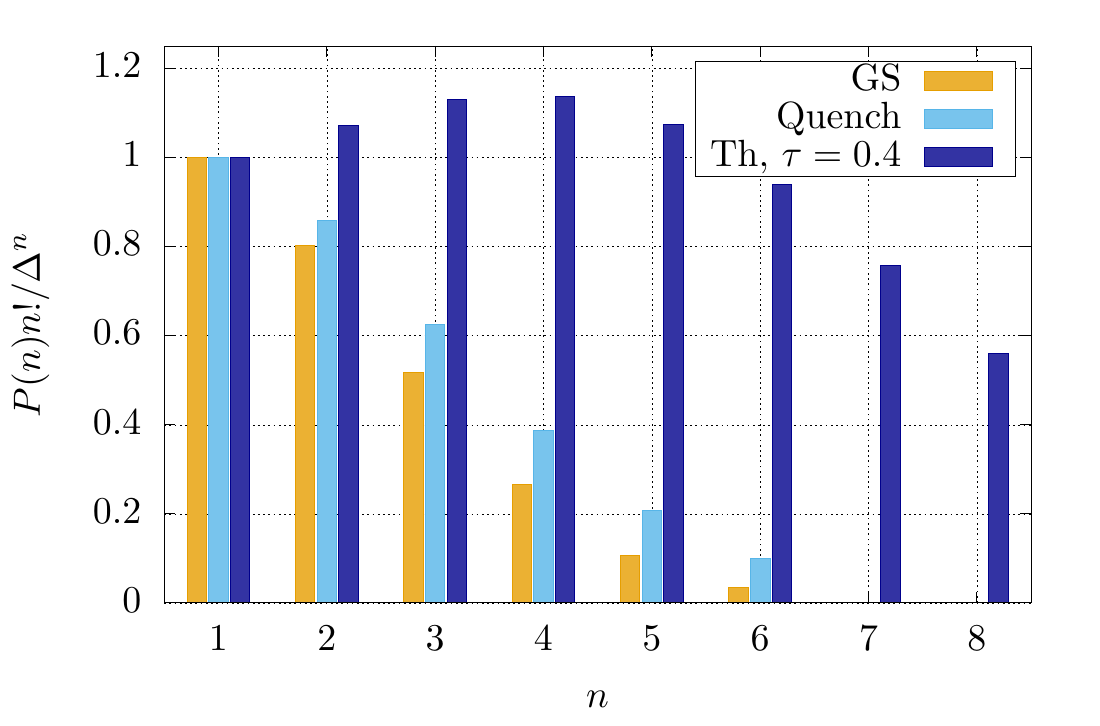}
		\caption{Rescaled probability distribution for particle-number fluctuations. The plot shows results for ground-state, thermal state with inverse temperature $\tau=0.4$, and the post-quench steady state studied in \cite{DWBC14}. The interaction is set to $\gamma=0.1$.}
		\label{fig:full_counting}
	\end{center}
\end{figure}
In order to establish~\eqref{eq:prob}, we consider the operator $N_\Delta= \int_{x-\Delta/2}^{x+\Delta/2}\dd y\,  \psi^\dagger(y)\psi(y)$, 
and introduce the generating function $\chi(\nu)=\langle e^{i\nu N_\Delta} \rangle$;  following \cite{KoCC14,BaCS17}, one can show
$ \chi(\nu)=\langle:e^{(e^{i\nu}-1)N_\Delta} :\rangle$,
where $:\ldots :$ denotes normal ordering. 
The Fourier transform of the generating function yields the full probability distribution corresponding to $N_\Delta$, namely
\be
\int\frac{\dd\nu}{2\pi} e^{-i\nu \mu}\chi(\nu)=\big\langle\delta \big(N_\Delta-\mu\big) \big\rangle=\sum_{n=0}^\infty \delta(n-\mu)P_\Delta(n)\,,
\ee 
yielding
\be
 P_\Delta(n)=\sum_{j=0}^\infty\frac{(-1)^{j}}{n! j!}  \Big\langle:\left(\int_{x-\Delta/2}^{x+\Delta/2} \dd y\, \psi^\dagger(y) \psi(y)\right)^{j+n} :\Big\rangle\, ,
\label{eq:P_N}
\ee
from which Eq.~\eqref{eq:prob} follows immediately. 
For finite intervals, Eq.~\eqref{eq:prob} approximates well \eqref{eq:P_N} if $\Delta$ is smaller than the inverse density, 
$\Delta\ll D^{-1}$, and than the scale of spatial variation of the correlators
$\Delta\ll\sqrt{D/\langle\partial_x\psi^\dagger(x)\partial_x\psi(x) \rangle}$. 
The probability distribution \eqref{eq:prob} for different macrostates is illustrated in Fig.~\ref{fig:full_counting}.

\emph{The Negro-Smirnov formula---} We finally sketch the idea behind the derivation of the main result \eqref{eq:main_corr}, while the most technical part of the computations will be presented elsewhere \cite{InPreparation}. The LL model can be obtained as the non-relativistic limit of the ShG field theory \cite{KoMT09} with action 
\be
\mathcal{S}_\text{ShG}=\int \dd x\dd t\,  \frac{1}{2}\partial_\mu\phi\partial^\mu\phi-\frac{c^4}{64\kappa }(\cosh(c^{-1}4\sqrt{\kappa}\phi)-1)\,,
\label{ShGaction}
\ee
where $\phi$ is a real scalar field, $c$ the speed of light and we set the bare mass to $1/2$. 
(see \cite{ckl-14,BaLM16} for generalizations to other relativistic field theories.) 
Also the ShG field theory admits a TBA description, with quasi-particle and hole rapidity distributions $\sigma(\theta)$, $\sigma_h(\theta)$ 
satisfying
\be\label{ShGbethe}
\sigma(\theta)+\sigma_h(\theta)=\frac{Mc\cosh\theta}{2\pi}+\int_{-\infty}^{\infty}\frac{{\rm d}\theta'}{2\pi}\,\varphi^\text{ShG}(\theta-\theta')\sigma(\theta')\,,
\ee
where $M$ is the renormalized mass, while the kernel reads
\be
\varphi^\text{ShG}(\theta)=\frac{\sin(\pi\alpha)}{\cos(\pi\alpha)-\cosh\theta}\,,
\ee
with $\alpha= 16\kappa/(8\pi c + 16 \kappa)$. 
As we have repeatedly stressed, vertex operators in the ShG model can be computed via the recently proposed formula 
by Negro and Smirnov \cite{NeSm13}. The latter was later simplified in \cite{BePC16}, and within our notations reads
\be
\frac{\langle e^{(k+1)c^{-1}4\sqrt{\kappa}\phi}\rangle}{\langle e^{kc^{-1}4\sqrt{\kappa}\phi}\rangle}=1+\frac{2\sin(\pi\alpha(2k+1))}{\pi}\int \dd\theta \, \bar{\vartheta}(\theta)e^\theta p_k(\theta)\, ,
\label{eq:BP}
\ee
where $\bar{\vartheta}$ is the filing function for  a  given state, while $p_k(\theta)$ is the solution of the integral equation
\be
p_k(\theta)=e^{-\theta}+\int \dd\theta' \vartheta(\theta')\chi_k(\theta-\theta')p_k(\theta'),
\ee
with
\be
\chi_k(\theta)=\frac{i}{2\pi}\left(\frac{e^{-i2k\alpha\pi}}{\sinh(\theta+i\pi\alpha)}-\frac{e^{i2k\alpha\pi}}{\sinh(\theta-i\pi\alpha)}\right)\, .
\ee

Our result for $g_n$ is derived by performing a non-trivial NR limit of Eq.~\eqref{eq:BP}. 
The mapping between the ShG and the LL models is probably best appreciated through the following mode-splitting \cite{KoMT09}
\be
\phi(t,x)=\left(e^{ic^2t/2}\psi^{\dagger}(t,x)+e^{-ic^2t/2}\psi(t,x)\right)\label{LL11}\, .
\ee
This mapping can be easily understood at the free point $\kappa=0$: once the relativistic field $\phi$ has been expressed in the momentum basis through the canonical relativistic mode-splitting, a low energy expansion provides the NR modes. The oscillating terms are simply the rest energies $E=\sqrt{m^2c^4+(\lambda c)^2}\simeq mc^2+\lambda^2/(2m)$ with $m=1/2$ and $\lambda$ the rapidity (e.g. momentum) in the NR model.

The mapping holds true also at the level of Bethe equations as well as thermodynamics. In fact, through the relation $\lambda=Mc \theta$, together with $\rho(\lambda) =(Mc)^{-1}\sigma(\theta)$ and $\rho_\text{h}(\lambda) =(Mc)^{-1}\sigma_\text{h}(\theta)$, the correspondence between Eq.\eqref{eq:thermo_bethe} and Eq. \eqref{ShGbethe} is readily derived. In this limit, the renormalized mass $M$ goes to the bare mass \cite{KoMT09}. 
Regarding the expectation values of the observables, the mode-splitting \eqref{LL11} leads to \cite{KoMT09}
\be\label{eq:op_NR}
\lim_\text{NR}\langle:\phi^{2n+1}:\rangle=0;\,\,\,\,\, \lim_\text{NR}\langle:\phi^{2n}:\rangle=\binom{2n}{n}\langle(\psi^\dagger)^n(\psi)^n\rangle\,.
\ee
Double dots here denote the normal ordered powers of the field. This relation is simply derived  dropping the fast oscillating phases in $\langle :\phi^n:\rangle$, after the mode expansion \eqref{LL11} has been employed.
The relation between a power of the renormalized field and its normal ordered counterpart is highly non-trivial and requires a careful 
analysis of the corresponding form factors \cite{KoMu93, KoMT09}. 
As a final highly technical but crucial ingredient, one can derive the following relation \cite{InPreparation}
\be
\lim_\text{NR}\langle e^{4q\sqrt{\kappa} \phi}\rangle= \lim_\text{NR}\Big\langle  :e^{2\sin(2q\kappa)\phi/\sqrt{\kappa}}:\Big\rangle\,.
\label{eq:normal_exp}
\ee
Next, we rescale $k=qc$ in the l.h.s. of Eq.~\eqref{eq:BP} and take the NR limit keeping $q$ constant; we obtain
\be
\frac{\langle e^{(q+c^{-1})4\sqrt{\kappa}\phi}\rangle}{\langle e^{q4\sqrt{\kappa}\phi}\rangle}=1+c^{-1}\partial_q\lim_{c\to\infty} \log\langle e^{4q\sqrt{\kappa}\phi}\rangle+...\, ,
\ee
where further orders in $\mathcal{O}(c^{-1})$ can be neglected. Comparing the above with the proper limit of the r.h.s. of  \eqref{eq:BP} and then integrating in $q$, we find an integral equation for $\lim_\text{NR}\langle e^{4q\sqrt{\kappa} \phi}\rangle$. Finally, a lengthy but straightforward expansion in $q$ of the expression for $\lim_\text{NR}\langle e^{4q\sqrt{\kappa} \phi}\rangle$, combined with eq. \eqref{eq:op_NR}-\eqref{eq:normal_exp}, provides the generating function \cite{InPreparation}
\be
\sum_{n=0}^{\infty}X^n\frac{D^n g_n}{(n!)^2(\kappa/2)^n}=\exp\left(\frac{1}{\pi\kappa}\sum_{n=1}^{+\infty} X^n \mathcal{B}_n\right)\,,
\label{eq:generating_function}
\ee
(cf. Eqs. (\ref{eq:int_1}) and (\ref{eq:int_2}) for ${\cal B}_n$), from which we  immediately derive 
our main result \eqref{eq:main_corr}.

\emph{Conclusions---} We have derived analytic expressions for the $n$-body local correlations in the LL model by taking the
non-relativistic limit of the ShG field theory. 
Our result \eqref{eq:main_corr} can be very efficiently evaluated and is valid for arbitrary states. 
It is straightforward to take into account spatial inhomogeneities arising from a confining potential, by means of local density approximation, as done for $g_3$ in \cite{KoCI11}.
These $n$-point functions lead to an analytic expression for the full counting statistics of the particle-number fluctuations in a 
small interval, which represents a unique exact analytic result for interacting integrable models.

Our study opens several interesting directions for future investigations. 
For example, there are other integrable field theories that in the non-relativistic limit give models relevant for cold atoms, 
such as the attractive LL model \cite{ckl-14} and gases with multiple species of fermions and bosons \cite{BaLM16}.
Unfortunately, in these cases, a generalization of the Negro-Smirnov formula \cite{NeSm13} for the one-point function of vertex operators is not yet known.
We hope that the results in this Letter will boost its derivation. 
Our results can also be used as starting point for the determination of correlation functions in inhomogeneous setting, 
e.g. on an Eulerian scale \cite{Doyon17}, 
or in the low energy regime using conformal field theories in curved backgrounds \cite{dsvc17,bd-17}.

\emph{Acknowledgments---}. We acknowledge helpful discussions with Bal\'azs Pozsgay and M\'arton Kormos.


\end{document}